\documentstyle[12pt,epsf,psfig]{article}
\begin{document}
\title{\bf  LFV decays and  anomalous magnetic (electric) moments  in a lepton mass
matrices ansatz induced by SUSY GUT }
  \author{ Wu-Jun Huo
 \\
{\small\it  Department of  Physics, Peking University, Beijing
$100871$, P.R. China}}
\date{}
\maketitle
\begin{abstract}
By using the anomalous magnetic and electric dipole moments of the
$\tau$ lepton in an effective lagrangian approach to the new
physics, we investigate the lepton flavor violation (LFV) decays,
$l\to l'\gamma$, and $\mu,\tau$ anomalous magnetic and electric
dipole moments in a lepton mass matrices ansatz which induced by
SUSY GUT. We put very stringent constraints LFV decays and $\tau$
anomalous magnetic and electric dipole moments.
\end{abstract}

 \vspace*{0.5cm} \noindent

\newpage

\section{Introduction}

The discovery of neutrino oscillation has been one of the most
exciting experimental results in the recent years\cite{first}. As
the experimental data accumulating in the Super-Kamiokande
collaboration \cite{new,sma} and more results published by
SNO\cite{sno}, K2K\cite{k2k} and CHOOZ\cite{chooz} experiments,
the parameter space for the neutrino masses and mixing is
narrowing down considerably. The recent analysis show that the
atmospheric neutrino oscillation favors the $\nu_\mu-\nu_\tau$
process with best fitted maximal mixing, $\sin^22\theta_{atm}=1$
\cite{atm}. Among the four solutions for the solar neutrino
deficits, the large mixing angle MSW (LMA) solution is most
favored with the best fit values $\tan^2\theta_{sol}=0.37$ and
$\Delta m^2_{sol}=3.7\times 10^{-5} eV^2$\cite{lma}, followed by
the LOW and VAC solutions\cite{lma}. The small mixing angle (SMA)
solution is ruled out at the 2$\sigma$ level\cite{sma,lma}.

However, a survey of the theoretical models in the literature shows that most
of them yield the SMA or VAC
solution for the solar neutrino problem\cite{barr}.
The reason is that we usually believe that the exotic features of neutrinos,
compared to that of the quarks,
be attributed to their Majarana properties. Thus it is
commonly to reconcile the two large mixing angles and a small
mass squared splitting ratio $r\equiv \Delta m_{sol}^2/\Delta
m_{atm}^2 \approx 1.4\times 10^{-2}$ in one neutrino mass matrix $M_\nu$.
However, it is proved that this is very difficult to achieve.

In ref. \cite{model}, the authors  have built a phenomenological
neutrino model, in which they can attribute both the maximal
atmospheric neutrino mixing and the large solar neutrino mixing to
the charged lepton mass matrix. This happens naturally if one
choose an asymmetric form of the charged
lepton mass matrix. 
The charged lepton and neutrino mass matrices are given by
\begin{equation}
\label{form}
M_L=\left( \begin{array}{ccc} 0 & \delta & \sigma \\ -\delta & 0 & 1-\epsilon\\
0 & \epsilon & 1 \end{array}\right) m,\ \ M_\nu=\left(
\begin{array}{ccc} m_1 & 0 & 0 \\ 0 & m_2 & 0\\ 0 & 0 & m_3
\end{array}\right)\ ,
\end{equation}
with $\sigma\sim {\cal O}(1)$, $\epsilon\ll 1, \delta\ll
\epsilon$. For simplicity one have assumed that $M_\nu$ is
diagonal here. Taking the values of the parameters
\begin{equation}
\delta=0.00079,\ \epsilon=0.12,\ {\rm and}\ \ \sigma=0.55
\end{equation}
one can  obtain the correct mass ratios $m_e/m_\mu$,
$m_\mu/m_\tau$ and the MNS matrix
\begin{equation}
\label{vmns}
U_{\rm MNS}=\left(\begin{array}{ccc} 0.851 & -0.525 & -0.0056 \\
    0.362 & 0.595 & -0.718 \\ -0.380 & -0.609 & -0.696 \end{array} \right)\ .
\end{equation} This model then predicts the neutrino mixing parameters as
\begin{equation} \sin^22\theta_{atm}=0.999,
\tan^2\theta_{sol}=0.38,\,\,\, {\rm and}\,\, U_{e3}=-0.0056 ,
\end{equation}
which is in excellent agreement with the present most favored values.

Motivated by the announcement of the improved result for $a_\mu$
by BNL $g_\mu-2$ Collaboration \cite{g-2}, we consider an
effective Lagrangian approach to investigate the implications of
the above model on $\delta a_\mu$, $\delta a_\tau$ and the LFV
processes, $\mu\to e\gamma$, $\tau\to \mu\gamma$ and $\tau\to
e\gamma$. For the muon anomalous magnetic moment, the revised
difference between experiment and SM is
 \begin{eqnarray}
\delta a^{\rm SM}_\mu \equiv a^{\rm exp}_\mu -a^{\rm SM}_\mu
=16\times 10^{-10}
\end{eqnarray}
which is  a $1.6\sigma$ deviation. It might seem that the absolute
magnitude of the deviation may be a hint of new physics.  Many of
these new physics effects can also contribute to $a_\tau$ and
lepton flavor violation processes, such as $\tau\to l\gamma$
($l=\mu, e$) and $\mu\to e\gamma$.  For $\tau$, experimentally,
the  recent analysis of the $e^+ e^- \rightarrow \tau^+ \tau^-
\gamma$ process
 give \cite{taylor}
\begin{eqnarray}
 a_\tau = 0.004 \pm 0.027 \pm 0.023,\,\,\,
d_\tau = ( 0.0 \pm 1.5 \pm 1.3) \times 10^{-16} {\it e}~{\rm cm},
\end{eqnarray}
compared with the theoretical prediction of SM \cite{sm}
\begin{eqnarray}
a_\tau = 1.1769(4) \times 10^{-3}
\end{eqnarray}
and a very tiny $d_\tau$ from CP violation in the quark sector. In
ref. \cite{escribano},
 the authors put more stringent bounds of $a_\tau$ and $d_\tau$, which inferred from $Z$
 wedth $\Gamma(Z\to\tau^+ \tau^-)$ at LEP,
 \begin{eqnarray}
 -0.004 \leq a_\tau \leq 0.006 ,\,\,\,
| d_\tau | \leq 1.1 \times 10^{-17} {\it e}~{\rm cm}.
\end{eqnarray}
However, the most stringent bounds were put in ref. \cite{huang}.
Huang, Lin and Zhang firstly introduced the non-universal magnetic
and electric moment operators to the effetive lagrangian in the
lepton part and used a Fritzsch-Xing lepton mass matrix ansatz  to
pushed the anomalous magnetic and electric dipole moments of
$\tau$ lepton down to \begin{eqnarray}
 |a_\tau | < 10^{-11} ,\,\,\,
| d_\tau | <10^{-25} {\it e}~{\rm cm}.
\end{eqnarray}
 These results were obtained only in
one Fritzsch-Xing ansatz \cite{xing1,xing27} which induce the
nearly bi-maximal mixing pattern for atmospheric and solar
neutrino oscillations \cite{altarelli}. In fact, there are several
other deferent lepton mass matrix ansatzs, such as our ansatz
above which  induced from SUSY GUT,  that are compatible with the
current neutrino oscillation experimental data.

 In Sec. 2, we give the general form of the effective
 lagrangian approach to new physics which was presented in Ref.
 \cite{huang}. The LFV decays, lepton anomalous magnetic and electric dipole
 moments in our
lepton mass matrices ansatz and numerical analysis will be given
in Sec. 3. We
 get our conclusion in the last section.

\section{ Effective Lagrangian with magnetic and electric moment operators}

As in refs. \cite{escribano,huang}, to study magnetic and electric
moments beyond SM, the authors considered
 an effective lagrangian approach to new physics,
\begin{eqnarray}\label{Leff}
{\cal L}_{\rm eff} =
{\cal L}_{\rm SM} + \frac{1}{\Lambda^2} \sum_i C_i {O}_i ,
\end{eqnarray}
where
 $\Lambda$ is the new physics
scale and $O_i$ are $SU_c(3)\times SU_L(2)\times U_Y(1)$ invariant
operators. $C_i$ are constants which represent the coupling
strengths of $O_i$. The complete operators list can be found in
Ref. \cite{young}. For the anomalous magnetic moment of the tau
lepton, there  have two dimension-six operators \cite{escribano}
\begin{eqnarray}\label{operator}
{O}_{\tau B}&=&{\bar L_\tau}\sigma^{\mu\nu} \tau_R \Phi B_{\mu\nu}, \\
{O}_{\tau W}&=&{\bar L_\tau} \sigma^{\mu\nu} {\vec \sigma} \tau_R \Phi {\vec
W}_{\mu\nu}.
\end{eqnarray}
Here $L_\tau $ is the $\tau$ left-handed isodoublet, $(
\nu_{\tau},  \tau_L )$, $\tau_R $ is right-handed singlet, $\Phi$
is the  scalar doublet, $B_{\mu\nu}$ and $W_{\mu\nu}$ are $U_Y(1)$
and $SU_L(2)$ field strengths. Similarly, operators below are
introduced to induce the electric dipole moments of $\tau$
\cite{escribano},
\begin{eqnarray}
\label{eq23}
\tilde{\cal O}_{\tau B}&=&{\bar L}\sigma^{\mu\nu}i\gamma_5 \tau_R \Phi B_{\mu\nu}, \\
\label{24} \tilde{\cal O}_{\tau W}&=&{\bar
L}\sigma^{\mu\nu}i\gamma_5{\vec \sigma} \tau_R \Phi
                   {\vec W}_{\mu\nu},
\end{eqnarray}

When $\Phi$ gets vacuum expectation value, operators $O_{\tau B}$
and $O_{\tau W}$ can give rise to anomalous magnetic moment of
lepton. After the broken of  electroweak symmetry and
diagonalization of  the mass matrices of the leptons and the gauge
bosons, the   effective neutral current couplings of the leptons
to the photon $\gamma$ are
\begin{eqnarray}\label{zgamma}
{\cal L}^{ \gamma}_{\rm eff}&=& {\cal L}^{ \gamma}_{\rm SM}
+eg^{\gamma}
                    \frac{1}{ 2 m_\tau}( -i k_\nu \sigma^{\mu\nu}) S^{ \gamma}
 {{ \pmatrix{{\overline e}
  \cr {\overline \mu} \cr
  {\overline
   \tau} \cr}}}^T U_l
 \pmatrix{ 0 & 0& 0\cr
  0& 0 &0 \cr
  0&0 &  1 \cr }
  U_l^\dagger
  \pmatrix{e \cr \mu \cr \tau \cr}, \nonumber \\
& &
\end{eqnarray}
where $ g^\gamma=1$, $V^{\gamma}=1-4s_W^2,1$, and
\begin{eqnarray}
\label{sr}
 S^{\gamma}&=&\frac{2m_\tau}{e}\frac{\sqrt 2 v}{\Lambda^2}\left [
   C_{\tau W }\frac{s_W}{2}-C_{\tau B }c_W \right ].
\end{eqnarray}
Matrix $U_l$ is the unitary matrix which diagonlizes the mass
matrix of the charged lepton. Here, we take the case $M_l^{\rm
diag} =U_l M_l U_l^\dagger$. In our ansatz, because the mass
matrix of neutrino is diagonlized, the matrices $U_l$ are really
the lepton mixing matrix $U_{\rm MNS}$. We can note
\begin{equation}\label{v1}
{V'}=U_{\rm MNS}\left (
\begin{array}{lcr}
0 & 0 & 0\\
0& 0 & 0\\
0 & 0 & 1
\end{array}
\right )U_{\rm MNS}^{\dagger}.
\end{equation}
From the above equation, the decay width of $l\rightarrow
l'+\gamma$ is given by
\begin{eqnarray}\label{ll}
\Gamma{(l\rightarrow l'\gamma)}=
\frac{m_l}{32\pi}\left({V'}_{ll'}e S^{\gamma}
\frac{m_l}{m_{\tau}}\right)^2,
\end{eqnarray}
where ${V'}_{ll^{\prime}}$ ($l\not=l^{\prime}$) are the
non-diagonal elements of matrix $V'$ defined in Eq. (17). The new
physics contribution to the anomalous magnetic moments of the
lepton is given by
\begin{eqnarray}\label{att}
\vert \delta \alpha_l \vert = \left \vert {V'}_{ll} S^{\gamma}
\right \vert.
\end{eqnarray}
From the above equations, after given the lepton mixing matrix
$U_{\rm MNS}$, we can obtain the  matrix $V'$ and put the
constraints of $\delta a_\tau$ from the LFV decays $l\to
l'\gamma$, and vice versa.

\section{The lepton mass matrices ansatz from SUSY GUT and the numerical results}

Experimentally, the observed neutrino oscillations indicate that
neutrinos are massive and lepton flavors are mixed. By putting
aside the LSND experiment and considering the current atmospheric
and solar neutrino oscillation data, in the framework of three
light neutrinos,  one can put the strong constraints of  the
lepton mixing matrix $U_{\rm MNS}$. Theoretically, There have been
many models which can predict lepton mixing matrices. Of course,
the models of lepton mass matrices $M_l$ and $M_\nu$ should
naturally lead to hierarchical neutrino mass squared differences
and large lepton flavor mixing angles which be required by current
neutrino oscillation data.

In ref. \cite{model}, there is  a phenomenological neutrino model,
in which has both the maximal atmospheric neutrino mixing and the
large solar neutrino mixing to the charged lepton mass matrix. we
can  obtain the MNS matrix $U_{\rm MNS}$ (Eq. (3)) which is in
excellent agreement with the present most favored values, and
matrix $V$
\begin{eqnarray}
{V'}=
 U_{\rm MNS} \left (
\begin{array}{lcr}
0 & 0 & 0\\
0& 0 & 0\\
0 & 0 & 1
\end{array}
\right )U_{\rm MNS}^{\dagger}=  \left (\matrix{ 0.00003 & 0.0038 &
0.0036 \cr 0.0038 & 0.5285 & 0.4994 \cr 0.0036 & 0.4994 & 0.472
\cr } \right ),
\end{eqnarray}

From Eqs. (18), (19) and the above matrix $V'$, given the current
experimental upper limits on $\mu\to e\gamma$ \cite{muer},
$BR(\mu\to e\gamma) \leq 1.2 \times 10^{-11}$,
we have \begin{eqnarray} |S^\gamma |&\leq& 1.4\times
10^{-10}\nonumber\\
|\delta \alpha_\tau |&\leq &1.1\times 10^{-11}
\end{eqnarray}
Similarly, from the effective operators in Eqs. (13) and (14), we
put limit on the anomalous electric dipole moment
\begin{eqnarray}
|d_\tau |\leq 2.7\times \times 10^{-25}.
\end{eqnarray}
We find these results are  are very stringent compared with their
experimental bounds.

Moreover, also from Eq. (18), we have
\begin{eqnarray}
\frac{\Gamma{(\mu \rightarrow e\gamma)}}{\Gamma{(\tau \rightarrow
l\gamma)}}= \left(\frac{m_\mu}{m_\tau}\right)^3
\left(\frac{|V'_{\mu e} |}{|V'_{\tau l} |}\right)^2.
\end{eqnarray}
By using the values of lifetime of $\mu $ and $\tau$, $\tau_\mu$
and $\tau_\tau$ \cite{muer}, we obtain
\begin{eqnarray}
BR (\tau\to \mu \gamma) &\leq& 1.6 \times 10^{-10}\\
BR (\tau\to e \gamma) &\leq& 8 \times 10^{-13}
\end{eqnarray}
These results are much smaller than the experimental bounds
\cite{muer}. We  put the very  stringent constraints of LFV decays
of $\tau$, which can be tested by the future experiments.

From Eq. (19) and $V'$, we  get that
\begin{eqnarray}
\frac{|\delta a_\mu|}{|\delta
a_\tau|}=\frac{V'_{\mu\mu}}{V'_{\tau\tau}}=1.12
\end{eqnarray}
By using the above value of $\delta a_\tau$, we have
\begin{eqnarray}
\delta a_\mu \leq 1.2 \times 10^{-11}
\end{eqnarray}
This result are much smaller than the current 1.6 $\sigma$
experimental value of $g-2$ and the previous value which was
2.6$\sigma$ deviation from the SM. This shows that, perhaps along
with the improvement of experimental precision, the value of
$\delta a_\mu$ will decrease and change much smaller. That is say,
the contribution to $g-2$ from new physics is very small.

 \section{Conclusion}

 By using the effective lagrangian to the new physics,
 we investigate the lepton flavor violation and anomalous magnetic and electric dipole
 moments in a lepton mass matrices ansatz induced by SUSY GUT.
We put much more stringent bounds of $\tau$ anomalous magnetic and
electric dipole moment than the current experimental results. This
is similar to the conclusion of the Ref. \cite{huang}. We also
study the lepton flavor violation decays, $\tau\to \mu\gamma$ and
$\tau\to e\gamma$ and put much more stringent bounds than the
current experimental values. We also find that muon anomalous
magnetic moments from new physics is consistent with and much
smaller than the current $g-2$ experimental outcome.

\section*{Acknowledgments}

The author thanks Prof. Tao Huang and Dr. Xiao-Jun Bi for useful
discussion and acknowledges supports from the Chinese Postdoctoral
Science Foundation and CAS K.C. Wong Postdoctoral Research Award
Fund.

\end{document}